\documentclass{iaus}
\usepackage{graphics}
\usepackage{amssymb}

\title[Stellar populations in the Magellanic Clouds]
{Stellar populations in the Magellanic Clouds: looking through the dust}

\author[De Marchi, Panagia \& Romaniello]  %%    
{Guido De Marchi$^1$, Nino Panagia$^2$\thanks{also INAF-HQ, Via del Parco 
 Mellini 84, 00136 Rome, Italy; and Supernova Ltd., OYV \#131, Northsound
 Road, Virgin Gorda, British Virgin Islands} \and Martino Romaniello$^3$}

\affiliation{
$^1$ESA, Space Science Department, Keplerlaan 1, 2200 AG 
Noordwijk, The Netherlands \break email: gdemarchi@rssd.esa.int\\[\affilskip]
$^2$Space Telescope Science Institute, 3700 San Martin Drive, Baltimore, 
MD 21218, USA \break email: panagia@stsci.edu\\[\affilskip]
$^3$European Southern Observatory, Karl-Schwarzschild-Strasse 2, 85748
Garching, Germany \break email: mromanie@eso.org}

\pubyear{2007}
\volume{241}  %% insert here IAU Symposium No.
\pagerange{1-2}
\date{?? and in revised form ??}
\jname{Stellar Populations as Building Blocks of Galaxies}
\editors{A. Vazdekis \& R. Peletier, eds.}
\begin{document}

\maketitle

\begin{abstract}
We present the first results of our study of stellar populations in the
Large and Small Magellanic Clouds based on multi-band WFPC2 observations
of ``random'' fields taken as part of the ``pure parallel'' programme
carried out with the HST as a service to the community.
\keywords{Hertzsprung-Russell diagram; stars: fundamental parameters; 
dust, extinction}
\end{abstract}

\firstsection % if your document starts with a section,
\section*{Summary}

We have started a study of the stellar populations in the Large and Small 
Magellanic Clouds using data (U, B, V, I and H$\alpha$ imaging) collected 
as part of the WFPC2 Pure Parallel Program (Wadadekar et al. 2006). We 
have considered a number of fields comprising both young and old stellar 
populations, with ages ranging from a few Myr to a few Gyr and where also
pre-main-sequence objects are clearly identified. In addition to the 
properties of these populations, we have studied the characteristics of 
the dust present in these regions. 

Using the properties of red giant stars of the so-called ``red clump''
(RC) in the various colour--magnitude diagrams, we have determined the 
extinction law in each field (Panagia \& De Marchi 2005; De Marchi \&
Panagia 2007). The RC is populated by stars experiencing core He burning 
and its position in the CMD depends on: (1) the age and (2) metallicity
of the population, (3) its distance and (4) the intervening absorption 
($A_\lambda$). Our high precision and high resolution photometry in four 
or more bands allows us to disentangle the four effects and determine or 
constrain their values. For fields with large and variable extinction, 
the RC is particularly useful to derive the extinction value towards each 
individual RC star, i.e. the extinction towards old populations.

We find that not only the amount of extinction but also its wavelength
dependence varies considerably from site to site, indicating that the
physical properties of the absorbing dust are not uniform across the
Clouds. For instance, we have determined the extinction law towards a 
region located $6^\prime$ SW of 30~Dor and derived the absolute 
value of $A_\lambda$ for all RC stars. The extinction law compares well 
with that obtained spectroscopically in a region containing SN 1987A 
(Scuderi et al. 1996), but it is significantly shallower in the U and B 
bands (see Figure\,1), thus implying the presence of relatively large 
dust grains. Analytically, our extinction law in that specific field is 
equivalent to the Galactic diffused ISM law ($R_V=3.1$) multiplied by a 
power law $\lambda^{0.5}$.

This analysis also allows us to determine the relative three-dimensional
distribution of the different populations as well as their location
with respect to the absorbing dust clouds. For instance, in the field 
near 30~Dor mentioned above we find that all young stars lie behind a 
layer of dust with $E(B-V) \simeq 0.3$ and very few of them extend past
$E(B-V) \simeq 0.7$, whereas the reddening towards RC stars spans the
full range $0 \lesssim E(B-V) \lesssim 1$, indicating that the latter 
stars are evenly distributed along the line of sight. Furthermore, 
although the reddening distribution that we derive for all RC stars is 
compatible with that of Zaritsky et al. (2004) for the same field, it 
spans a considerably wider range of reddening.

Correcting for the appropriate extinction for each field {\it
individually}, and following  the procedures developed by Panagia et al.
(2000), Romaniello et al. (2002), and Romaniello, Robberto \& Panagia
(2004), we determine the physical parameters, namely effective
temperature and luminosity, of all stars present in the field (De Marchi
et al. 2005). We find that, among young stars, those less massive than 
2\,M$_\odot$ are spatially more spread over each field than massive B type
stars, whereas old stars are quite uniformly distributed. All young stars
appear to be spatially associated with the nebular emission.

An important corollary of our investigation is that unaccounted patchy
absorption or a variable extinction law are likely to contribute
significantly to the present discrepancies between various distance
indicators for the Magellanic Clouds. 

\begin{figure}
\centering
\resizebox{13cm}{7cm}{\includegraphics{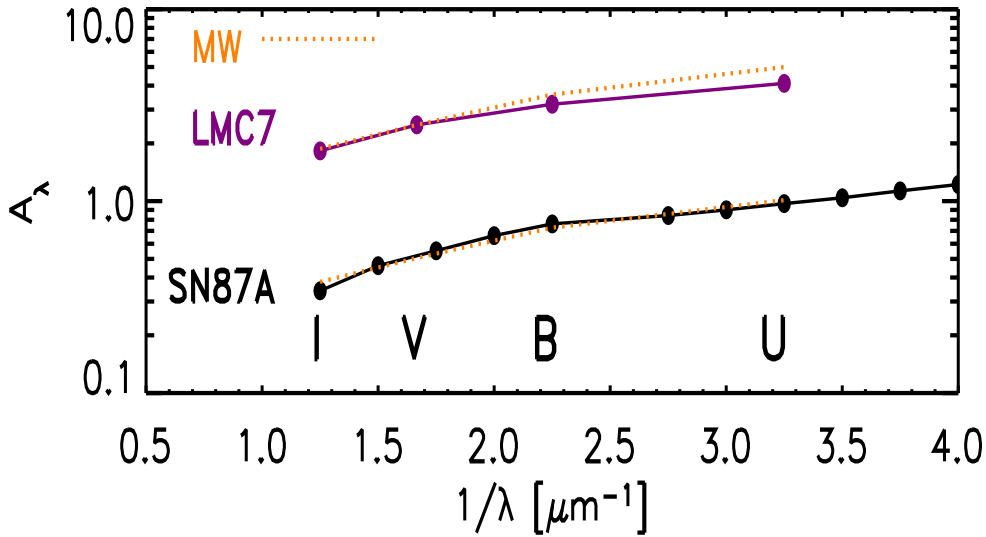} }
\caption{Photometrically determined extinction law in a field
  $6^\prime$ SW of 30~Dor (LMC7) is compared with that of the
  field containing SN1987A and to the Galactic law (dotted line).}
\end{figure}

%\begin{discussion}

%\end{discussion}

\end{document}